\documentclass[journal,comsoc]{IEEEtran}

\ifCLASSINFOpdf
   \usepackage[pdftex]{graphicx}
\else

\fi

\ifCLASSOPTIONcomsoc
	\usepackage{amsmath}
	\usepackage[cmintegrals]{newtxmath}
	\usepackage{bm} 
\else
	\usepackage{amsmath}
	\usepackage{bm}
\fi

\ifCLASSOPTIONcompsoc
  \usepackage[caption=false,font=normalsize,labelfont=sf,textfont=sf]{subfig}
\else
  \usepackage[caption=false,font=footnotesize]{subfig}
\fi

\usepackage{mleftright}
\usepackage{pgfplots}
\usepackage{ wasysym }
\usepackage{trfsigns}

\usepackage{siunitx}
\usepackage{booktabs}
\usepackage[normalem]{ulem}

\usepackage[nolist]{acronym}
\acrodef{AWGN}[AWGN]{additive white Gaussian noise}

\acrodef{BPS}[BPS]{blind phase search}

\acrodef{CD}[CD]{chromatic dispersion}
\acrodef{CMA}[CMA]{constant-modulus algorithm}
\acrodef{CNN}[CNN]{convolutional neural network}
\acrodef{CPE}[CPE]{carrier-phase estimation}

\acrodef{DFE}[DFE]{decision feedback equalizer}
\acrodef{DP}[DP]{dual-polarization}
\acrodef{DSP}[DSP]{digital signal processing}

\acrodef{ELU}[ELU]{exponential linear unit}
\acrodef{ELBO}[ELBO]{evidence lower bound}

\acrodef{FEC}[FEC]{forward error correction}
\acrodef{FIR}[FIR]{finite impulse response}

\acrodef{IP}[IR]{impulse response}
\acrodef{ISI}[ISI]{inter-symbol interference}

\acrodef{KL}[KL]{Kullback-Leibler}
\acrodef{LDPC}[LDPC]{low-density parity-check}

\acrodef{MA}[MA]{moving average}
\acrodef{MAP}[MAP]{maximum a posteriori}
\acrodef{MIMO}[MIMO]{multiple-input multiple-output}
\acrodef{ML}[ML]{maximum likelihood}
\acrodef{MMA}[MMA]{multi-modulus algorithm}
\acrodef{MMSE}[MMSE]{minimum mean squared error}

\acrodef{NN}[NN]{neural network}

\acrodef{PCS}[PCS]{probabilistic constellation shaping}
\acrodef{pdf}[pdf]{probability density function}
\acrodef{PMD}[PMD]{polarization mode dispersion}
\acrodef{pmf}[pmf]{probability mass function}
\acrodef{PSK}[PSK]{phase shift keying}
\acrodef{PSP}[PSP]{principal state of polarization}

\acrodef{QAM}[QAM]{quadrature amplitude modulation}

\acrodef{RRC}[RRC]{root-raised cosine}

\acrodef{SER}[SER]{symbol error rate}
\acrodef{SNR}[SNR]{signal-to-noise ratio}
\acrodef{sps}[sps]{samples per symbol}

\acrodef{VAE}[VAE]{variational autoencoder}

\newcommand{\lr}[1]{\left(#1\right)}
\newcommand{\mlr}[1]{\mleft(#1\mright)}
\newcommand{\mlrb}[1]{\mleft\{#1\mright\}}

\definecolor{ALUColor1}{RGB}{77,175,74}
\definecolor{ALUColor2}{RGB}{228,26,28} 
\definecolor{ALUColor3}{RGB}{77,175,74}
\definecolor{ALUColor4}{RGB}{55,126,184} 
\definecolor{ALUColor5}{RGB}{255,127,0}
\definecolor{ALUColor6}{RGB}{152,78,163}
\definecolor{ALUColor7}{rgb}{0.6350,0.0780,0.1840}

\def\j{{\mathrm{j}}}

\def\real{{\mathrm{Re}}}

\def\I{{\mathrm{I}}}
\def\Q{{\mathrm{Q}}}

\def\TE{{\mathrm{TE}}}
\def\TM{{\mathrm{TM}}}

\def\H{{\mathrm{H}}}

\newcommand{\fft}[1]{\mathcal{F}\mleft\{#1\mright\}}

\DeclareMathOperator{\E}{\mathbb{E}}
\newcommand{\qxy}{Q_{\bm{\Phi}}\mlr{\bm{x}| \bm{y}}}
\newcommand{\elr}{\epsilon_{\mathrm{lr}}}

\newcommand{\sigw}{\sigma_{\mathrm{w}}}

\begin{document}
\title{Blind Equalization and Channel Estimation in Coherent Optical Communications Using Variational Autoencoders}

\author{Vincent~Lauinger,~\IEEEmembership{Student~Member,~IEEE}, Fred~Buchali~and~Laurent~Schmalen,~\IEEEmembership{Senior~Member,~IEEE}%
\thanks{This work was carried out in the framework of the CELTIC-NEXT project AI-NET-ANTILLAS (C2019/3-3) and was funded by the German Federal Ministry of Education and Research (BMBF) under grant agreement 16KIS1316.}%
\thanks{V.~Lauinger and L.~Schmalen are with Communications Engineering Lab (CEL), Karlsruhe Institute of Technology (KIT), 76131 Karlsruhe, Germany (email: vincent.lauinger@kit.edu, laurent.schmalen@kit.edu).}%
\thanks{F.~Buchali is with Nokia, 70469 Stuttgart, Germany (email: Fred.Buchali@nokia.com).}%
\thanks{Parts of this paper have been presented at the Advanced Photonics Congress, Signal Processing in Photonic Communications (SPPCom), 2021 \cite{lauinger2021blind}.}}

\markboth{IEEE Journal on Selected Areas in Communications, \textit{accepted for publication}, July~2022}%
{Lauinger \MakeLowercase{\textit{et al.}}: Blind Equalization and Channel Estimation in Coherent Optical Communications Using Variational Autoencoders}

\maketitle

\begin{abstract}
    We investigate the potential of adaptive blind equalizers based on variational inference for carrier recovery in optical communications. These equalizers are based on a low-complexity approximation of maximum likelihood channel estimation. We generalize the concept of \ac{VAE} equalizers to higher order modulation formats encompassing \ac{PCS}, ubiquitous in optical communications, oversampling at the receiver, and dual-polarization transmission. Besides black-box equalizers based on convolutional neural networks, we propose a model-based equalizer based on a linear butterfly filter and train the filter coefficients using the variational inference paradigm. As a byproduct, the \ac{VAE} also provides a reliable channel estimation.
    We analyze the \ac{VAE} in terms of  performance and flexibility over a classical \ac{AWGN} channel with \ac{ISI} and over a dispersive linear optical  dual-polarization channel. We show that it can extend the application range of blind adaptive equalizers by outperforming the state-of-the-art \ac{CMA} for \ac{PCS} for both fixed but also time-varying channels. The evaluation is accompanied with a hyperparameter analysis.
\end{abstract}


\acresetall

\section{Introduction}
	\IEEEPARstart{T}{he} digital transformation along with the modern lifestyle and the advent of video streaming platforms brought up a strong demand for high-speed and highly flexible communication systems. 
Precisely, the required data rates can only be provided by coherent optical communication systems along with high-order modulation formats and \ac{PCS}~\cite{bocherer2015bandwidth}. Due to its properties, e.g., easy rate adaption~\cite{BuchaliJLT}, a decreased gap to the \ac{AWGN} channel capacity, increased energy efficiency~\cite{cho2019probabilistic} and a larger tolerance against fiber non-linearities, \ac{PCS} has become an essential ingredient of modern coherent optical communication systems~\cite{Sun:20, buchali2020cmos}.
However, the use of \ac{PCS} entails a more challenging carrier recovery than conventional square \ac{QAM} formats. 
Often, data-aided or pilot-based algorithms are the only option nowadays, since an open issue in communications is the lack of optimum (but practical) blind adaptive channel equalizers.
However, pilot symbols cannot transport information so they reduce data rate and limit the achievable net bit rate significantly.
Hence, there is a strong need of blind channel equalizers, which can adapt to time-varying channels and transmission parameters. The saved data rate can be used to either increase the throughput or \ac{FEC} overhead.

In coherent optical communications, the standard algorithm for blind adaptive equalization of linear channels and symmetric complex modulation formats is the \ac{CMA} \cite{godard}. It tries to reach a constant signal amplitude (radius) by adaptively equalizing the signal with trainable \ac{FIR} filters. Thus, it is optimal for constant-amplitude formats such as $M$-ary \ac{PSK}, but it also converges for multi-amplitude formats such as $M$-ary \ac{QAM} \cite{johnson1998blind} where its criterion is sub-optimal. The \ac{MMA} \cite{yang2002multimodulus} is an extension for multi-amplitude formats, however, it suffers from its high implementation complexity and low convergence rate.
Based on the same criterion, a non-linear blind \ac{NN} based equalization scheme was proposed in \cite{you1998nonlinear}. 
Since the \ac{CMA}’s criterion is independent of the signal phase, detection is only possible in combination with a \ac{CPE} block. The commonly used algorithms are the blind phase search \cite{pfau2009hardware} or the Viterbi-Viterbi algorithm \cite{viterbi1983nonlinear}, which both face performance degradation for \ac{PCS} \cite{mello2018interplay, zhang2020viterbi}. Similarly, the \ac{CMA} suffers from convergence issues for \ac{PCS} as well \cite{chan1990stationary, zervas1991effects}.

Optimally, we want to use the \ac{ML} criterion, which has been considered for blind equalization, e.g., in \cite{ghosh1992maximum, tong1998multichannel, kaleh1994joint}.
However, we are not aware of any blind \ac{ML} based channel equalizer which has been seriously considered in real coherent optical communication systems. A promising approach is to approximate \ac{ML} by \emph{variational inference} via a \ac{VAE} \cite{kingma2013auto, blei2017variational}. 
Variational inference is used for unsupervised and semi-supervised learning as well as generative models, however, there are not many applications in communications with notable exceptions 
being~\cite{VI_E2E_raj2020design, caciularu2018blind, caciularu2020unsupervised}. While~\cite{VI_E2E_raj2020design} trains end-to-end transmission systems without \ac{ISI} in a supervised manner, we, in contrast, focus on blind \ac{VAE}-based equalization where we use unsupervised learning at the receiver to adjust the equalizer weights. Such an unsupervised equalizer implementation has initially been
presented in~\cite{caciularu2018blind} and further extended in~\cite{caciularu2020unsupervised} towards unsupervised \ac{LDPC} decoding. However, only a simple quadrature \ac{PSK} (QPSK) implementation was used. 
In this work, we generalize the approach of 
\cite{caciularu2018blind,caciularu2020unsupervised}
and propose essential extensions, including the application to oversampled \ac{DP} signals and multi-amplitude \ac{PCS} formats. 
We show that the generalized approach is independent of the equalizer architecture and train both a \ac{CNN} based equalizer as in \cite{caciularu2018blind, caciularu2020unsupervised}---the VAE-NN---and a novel linear model-based equalizer with butterfly \ac{FIR} filters---the VAE-LE. We evaluate the performance of the proposed equalizers on different linear channels and propose an extension for slowly time-varying channels.

This paper is structured as follows: in Sec.~\ref{sec:sys_model}, we introduce our system model, in Sec.~\ref{sec:VI}, we motivate variational inference for equalization, derive the \ac{VAE}-based equalizer in a general form and explain how the proposed extensions can be incorporated. In Sec.~\ref{sec:realization}, we discuss the implementation of the equalizers and propose an appropriate parameter update scheme, before we introduce our simulation environment and our results in Sec.~\ref{sec:results}. We conclude the paper in Sec.~\ref{sec:conclusion}.

\section{System Model} \label{sec:sys_model}

We start with the demonstration of the basic concept on a simple \ac{AWGN} channel with \ac{ISI}
\begin{align*}
	\bm{y} = \bm{h}_{\mathrm{sim}}\ast \bm{x} + \bm{n} \ ,
\end{align*}
where the transmit vector $\bm{x}$ is convolved with the simulated channel \ac{IP} $\bm{h}_{\mathrm{sim}}$ and a white noise vector $\bm{n}$ is added.

Then, we focus on a dispersive linear optical dual-polarization transmission to prove the concept in a practical environment (coherent optical transmission). We use the more natural description in the frequency domain by a linear 
channel matrix, i.e.,
\begin{align*}
	\begin{bmatrix} \bar{\bm{y}}_\TE \\ \bar{\bm{y}}_\TM \end{bmatrix} = \bm{H}\mlr{f} \cdot \begin{bmatrix} \bar{\bm{x}}_\TE \\ \bar{\bm{x}}_\TM \end{bmatrix} + \begin{bmatrix} \bar{\bm{n}}_\TE \\ \bar{\bm{n}}_\TM \end{bmatrix} \ ,
\end{align*}
with $\bar{\bm{z}} = \fft{\bm{z}}$ being the Fourier transform of a time domain vector $\bm{z}$ and $\{\TE, \TM\}$ being indices describing the two polarizations of the light. 
Additionally, we emulate a slowly time-varying channel by changing the channel matrix' elements over time, so we can analyze the performance in a dynamical environment.

Per se, the optical channel is nonlinear, but the linear distortions are dominant in practical systems and have to be compensated, whereas the nonlinearity compensation is computationally demanding and usually provides a \ac{SNR} gain of less than \SI{1}{dB}~\cite[Sec. 6.9.3]{Springer_handbook}.
Hence, we focus in this paper on linear impairments and assumes potential nonlinearities to be either negligible or compensated by a separate \ac{DSP} block, e.g., based on digital backpropagation, which can be switched on if required.

Further details on the simulation model including the specific parameters are provided later in Sec.~\ref{sec:results}.

\section{Variational Inference for Equalization} \label{sec:VI}
	The goal of communications is to transmit data to a receiver, which has to fully recover the information without knowledge of the actually transmitted data. This can be interpreted as an inference problem, where the received samples $\bm{y} \in \mathbb{C}^N$ are observed variables while the transmitted samples $\bm{x}\in \mathbb{C}^N$ are unobservable latent variables. The optimum decision is based on the maximum of the a posteriori probability distribution \cite[Ch.~4.1]{proakis2008digital}
\begin{align*}
	P\mlr{\bm{x}|\bm{y}} &= \frac{p\mlr{\bm{y}|\bm{x}} \cdot P\mlr{\bm{x}}}{p\mlr{\bm{y}}} \ ,
\end{align*}
where $p\mlr{\bm{y}|\bm{x}}$ is the likelihood of $\bm{y}$ given $\bm{x}$, $P\mlr{\bm{x}}$ is the prior probability and $p\mlr{\bm{y}}$ is the observations' marginal density (also called the \emph{evidence}). Throughout the paper, we denote \acp{pmf} by a capital $P\mlr{\cdot}$ and continuous densities by a lower case $p\mlr{\cdot}$. While the prior and the likelihood can usually be assumed as known or modeled well, the evidence is commonly intractable to compute, since the marginalization's complexity grows exponentially with the length and symbol order of $\bm{x}$, i.e. $p\mlr{\bm{y}}= \sum_{\bm{x}} P\mlr{\bm{x}} p\mlr{\bm{y}|\bm{x}}$. \\

In statistics, this is a common problem which can be solved by \emph{variational inference}. It is also used in machine learning when the conditional has to be approximated efficiently and reliably \cite{blei2017variational}. In particular, this is the case in our problem where we usually require fast convergence to cope with time-dependent distortions.
The main idea is to cast inference into an optimization problem, where the goal is to find an approximation $Q\mlr{\bm{x}|\bm{y}} \in \mathcal{Q}$ of the true a posteriori \ac{pmf} $P\mlr{\bm{x}|\bm{y}}$ from a family of approximate \acp{pmf} $\mathcal{Q}$, parameterized by free \emph{variational parameters} \cite{blei2017variational}, over the latent variables.

\subsection{The Evidence Lower Bound (ELBO)}
A suitable objective function is the relative entropy $\textrm{D}_\textrm{KL}\mlr{P\|Q}$, also called the \ac{KL} divergence, which is an information-theoretical measure of proximity. It is asymmetric, non-negative and convex with its minimum at $Q = P$ \cite[Ch.~2]{cover2006elements}. Then, the optimization's goal is to find the best approximation to the true a posteriori probability for the observed varibles $\bm{y}$ by
\begin{align*}
	\hat{Q}\mlr{\bm{x}|\bm{y}} = \arg\min_{Q\in \mathcal{Q}} \textrm{D}_{\textrm{KL}}\mlr{Q\mlr{\bm{x}|\bm{y}}\|P\mlr{\bm{x}|\bm{y}}} \ .
\end{align*}
With $\E_Q\mleft\{\cdot\mright\} = \E_{Q\mlr{\bm{x}|\bm{y}}}\mleft\{\cdot\mright\}$ being the expectation regarding the variational approximation, the \ac{KL} divergence can be expressed as\footnote{Note that we use the natural logarithm with base e ($\ln$) instead of the logarithm with base 2 ($\log_2$), which is frequently used in communications and information theory.}
\begin{align}
	\textrm{D}_{\textrm{KL}}\mlr{Q\mlr{\bm{x}|\bm{y}}\|P\mlr{\bm{x}|\bm{y}}} &= \E_{Q}\mleft\{\ln{Q\mlr{\bm{x}|\bm{y}}}\mright\} - \E_{Q}\mleft\{\ln{P\mlr{\bm{x}|\bm{y}}}\mright\} \nonumber \\ 
	&= \E_{Q}\mleft\{\ln{Q\mlr{\bm{x}|\bm{y}}}\mright\} - \E_{Q}\mleft\{\ln{p\mlr{\bm{y}|\bm{x}}}\mright\} \nonumber \\ 
	&\hspace*{2.2ex} -\E_{Q}\mleft\{\ln{P\mlr{\bm{x}}}\mright\}
	+ \ln{p\mlr{\bm{y}}}\ . \label{eq:KL}
\end{align}
Since the \ac{KL} divergence depends on $\ln{p\mlr{\bm{y}}}$, it is not easily computable and thus not suitable as objective function. However, the evidence is independent of $Q\mlr{\bm{x}|\bm{y}}$, so the last term in \eqref{eq:KL} is only an additive constant regarding the optimization. Hence, maximizing the \emph{evidence lower bound}
\begin{align}
	\mathrm{ELBO}\mlr{Q} &= \E_{Q}\mleft\{\ln{p\mlr{\bm{y}|\bm{x}}}\mright\} +\E_{Q}\mleft\{\ln{P\mlr{\bm{x}}}\mright\} -\E_{Q}\mleft\{\ln{Q\mlr{\bm{x}|\bm{y}}}\mright\} \nonumber \\
	&=\underbrace{ \E_{Q}\mleft\{\ln{p\mlr{\bm{y}|\bm{x}}}\mright\} }_{=:\textrm{B}} - \underbrace{ \textrm{D}_\textrm{KL}\mlr{Q\mlr{\bm{x}|\bm{y}}\|P\mlr{\bm{x}}} }_{=:\textrm{A}} \label{eq:ELBO}
\end{align}
is equivalent to minimizing \eqref{eq:KL}. This mirrors the usual balance between likelihood and prior, since the ELBO's first term B, the expected likelihood, favors densities which explain the observed data, while the second term A encourages the densities to be close to the prior \cite{blei2017variational}.
The complexity of this optimization is defined by the complexity of the family $\mathcal{Q}$. 

By rearranging \eqref{eq:KL} and due to the \ac{KL} divergence's non-negativity, we can show that the ELBO lower-bounds the (log) evidence, i.e.,
\begin{align*}
	\ln{p\mlr{\bm{y}}} &= \textrm{D}_\textrm{KL}\mlr{Q\mlr{\bm{x}|\bm{y}}\|P\mlr{\bm{x}|\bm{y}}} + \mathrm{ELBO}\mlr{Q} \\
	&\geq \mathrm{ELBO}\mlr{Q} \ .
\end{align*}
Since $\ln{p\mlr{\bm{y}}}$ is a fixed upper bound, we sandwich $\textrm{D}_\textrm{KL}\mlr{Q\mlr{\bm{x}|\bm{y}}\|P\mlr{\bm{x}|\bm{y}}}$ by maximizing the ELBO, so we eventually minimize the \ac{KL} divergence and find a good approximation $Q\mlr{\bm{x}|\bm{y}}$ for the true a posteriori probability.

The concept can also be interpreted from a communication theory perspective with the likelihood $p\mlr{\bm{y}|\bm{x}}$ as a probabilistic \emph{encoder} and the a posteriori---or its variational approximation $Q\mlr{\bm{x}|\bm{y}}$, respectively---as the corresponding \emph{decoder}.\footnote{In the deep learning literature, e.g.,~\cite{kingma2013auto}, an opposite definition is often found where $p\mlr{\bm{y}|\bm{x}}$ is the decoder and $Q\mlr{\bm{x}|\bm{y}}$ is the encoder. However, from a communications  point-of-view we find our definition more intuitive.} Precisely, during transmission, the data $\bm{x}$ (latent variables) is encoded into the observable receive samples $\bm{y}$, while the receiver tries to decode the transmitted data again, estimating $\bm{x}$ from $\bm{y}$. Assuming that the densities come from families of parameteric distributions, the concept can be implemented as a \acfi{VAE} using machine learning techniques, where typically both encoder and decoder are implemented as \acp{NN} \cite{kingma2013auto, caciularu2020unsupervised}.
However, if we have a suitable model, e.g., of the encoder $p\mlr{\bm{y}|\bm{x}}$, we do not have to apply an \ac{NN} to learn it but we can use the model directly, as done in the following subsection.

\subsection{The Variational Autoencoder (VAE)-based Equalizer} \label{sec:VAE}
	\begin{figure} [!tb] 
	\begin{center}
		\includegraphics{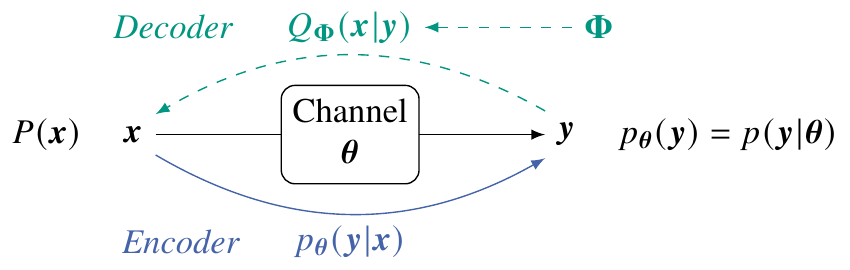}
		\caption{Simple transmission system and its corresponding \acp{pdf}: $\bm{x}$ and $\bm{y}$ are realizations of random variable sequences $\lbrace X_k\rbrace$ and $\lbrace Y_k\rbrace$;  $\bm{\theta}$ is a discrete-time \ac{AWGN} channel with impulse response $\bm{h}$ and noise variance $\sigma_{\mathrm{w}}^2$; $\qxy$ is an approximation (in the parameter space $\bm{\Phi}$) of the real a posteriori \ac{pdf}.}
		\label{VAE:simple_model}
	\end{center}
\end{figure}
We assume a general transmission system through a (parameterized) channel $\bm{\theta}$, as depicted in Fig.~\ref{VAE:simple_model}.
Then, the evidence $p_{\bm{\theta}}\mlr{\bm{y}}$ as well as the likelihood $p_{\bm{\theta}}\mlr{\bm{y}| \bm{x}}$ are parameterized by $\bm{\theta}$, while the variational approximation $\qxy$ can be parameterized by a set of learnable parameters $\bm{\Phi}$, as denoted by the corresponding subscripts.

In other words, the channel (respectively the encoder) distorts the transmitted signal, while the decoder tries to find the mapping from the distorted received samples to the transmitted data. Thus, finding the optimum variational approximation corresponds to finding the optimized equalizer for this channel. Furthermore, $\qxy$ gives a soft-decision on the received symbols, so the \ac{VAE}-based equalizer also approximates an \ac{ML} receiver \cite{caciularu2020unsupervised}. 

Note that the exact values of $\bm{\theta}$ are unknown, so the proposed equalization concept is blind and the model parameters are learned simultaneously with the decoder. In fact, the evidence $p_{\bm{\theta}}\mlr{\bm{y}} = p\mlr{\bm{y}|\bm{\theta}}$ can also be interpreted as the likelihood regarding the channel parameters, so variational inference also approximates maximum likelihood channel estimation. This byproduct can be used, e.g., for joint communication and sensing. See, e.g., \cite{dorize2020capturing} for an example of capturing acoustic signals based on the channel \ac{IP}.\\

In the following, we derive the \ac{VAE}-based equalizer in a generalized form compared to \cite{caciularu2020unsupervised}, where it is only derived for a toy model with QPSK. We try to keep repetitions as short as possible, but it is unavoidable at some points to highlight the generalizations we did.
We start by assuming transmission over an \ac{AWGN} channel parameterized by $\bm{\theta} = \lr{\bm{h},\sigma_{\mathrm{w}}^2}$ with finite \ac{IP} $\bm{h}$ and noise variance $\sigma_{\mathrm{w}}^2$. We can model the likelihood as
\begin{align}
	p_{\theta}\mlr{\bm{y}| \bm{x}} &\sim \mathcal{CN}\mlr{ \bm{h}\ast\bm{x} , \sigma_{\mathrm{w}}^2 \bm{I}_N } \ . \label{eq:pyx_model}
\end{align}
Further, we consider transmission of independently modulated square-$M$-\ac{QAM} symbols $x_i$, so $\bm{x} = \lr{x_1,\, \ldots x_N} = \bm{x}^{\mathrm{I}} + \j \bm{x}^{\mathrm{Q}}$ is a vector of $N$ complex-valued symbols. Assuming further that I and Q have been modulated independently, than $x_i^{\mathrm{I}},\ x_i^{\mathrm{Q}} \in \mathcal{A} = \{A_1, \ \ldots\ , \ A_{\sqrt{M}}\}$ are conditionally independent given $\bm{y}$. Consequently, we can model 
\begin{align}
	\qxy &= \prod_{i=1}^N Q_{\bm{\Phi}}\mlr{x_i^{\mathrm{I}}| \bm{y}} \cdot Q_{\bm{\Phi}}\mlr{x_i^{\mathrm{Q}}| \bm{y}} \label{eq:qxy_model}
\end{align}
and define a vector $\bm{q}^{c,A_m}\mlr{\bm{y}}$, $c\in\{\mathrm{I},\mathrm{Q}\}$, containing the parametric functions evaluated for each symbol $\bm{q}^{c,A_m}\mlr{\bm{y}} := \lr{Q_{\bm{\Phi}}\mlr{x_1^c = A_m| \bm{y}}, \, \ldots Q_{\bm{\Phi}}\mlr{x_N^c = A_m| \bm{y}}}$, which only depend on $\bm{y}$ and $\bm{\Phi}$. Although we have a similar decoder model as \cite{caciularu2020unsupervised}, we consider multi-level signals and, thus, cannot simplify further by exploiting the normalization of probabilities. 

Then, similarly to \cite{caciularu2020unsupervised}, we create a minimization problem by defining the loss function $\mathcal{L}\mlr{\bm{\theta},\bm{\Phi},\bm{y}} := - \mathrm{ELBO}\mlr{Q} = \textrm{A} - \textrm{B}$ (see \eqref{eq:ELBO}), which only depends on both parameter spaces, $\bm{\theta}$ and $\bm{\Phi}$, as well as the received samples $\bm{y}$. 

The first term A can be easily computed by the standard formula of the KL divergence, i.e.,
\begin{align}
	\textrm{A}\mlr{\bm{\Phi}} &= \textrm{D}_\textrm{KL} \mlr{ \qxy \| P\mlr{\bm{x}}}\nonumber \\
	&= \sum_{i=1}^N \sum_{m=1}^{\sqrt{M}} \sum_{c\in\{\mathrm{I},\mathrm{Q}\}} \left[ q_i^{c,A_m}\mlr{\bm{y}} \cdot \ln{\frac{q_i^{c,A_m}\mlr{\bm{y}}}{P\mlr{x^c_i=A_m} }} \right] \ . \label{eq:A}
\end{align}
Since the term A does not break down to the entropy as in \cite{caciularu2020unsupervised} (due to the assumption of a uniform prior \ac{pmf}), an important feature is the inclusion of the prior density $P\mlr{\bm{x}}$ into the loss function, which implies the adaption to, e.g., \ac{PCS} \cite{bocherer2015bandwidth, BuchaliJLT}. Since state-of-the-art blind equalizers struggle with non-uniform priors \cite{chan1990stationary, zervas1991effects, arikawa2020wide}, simple inclusion is one of the major benefits of this concept.

The second term B can be re-written, similarly to \cite{caciularu2020unsupervised}, as
\begin{align}
	\textrm{B}\mlr{\bm{\theta},\bm{\Phi},\bm{y}} &= \E_{Q}\mlrb{\ln p_{\theta}\mlr{\bm{y}| \bm{x}}} \nonumber \\
	&= -N\ln\mlr{\mathrm{\pi}\sigma_{\mathrm{w}}^2} - \frac{1}{\sigma_{\mathrm{w}}^2} \cdot \underbrace{\E_{Q}\mlrb{\|\bm{y}-\bm{h}\ast\bm{x}\|^2 }}_{\textrm{C}} \ , \label{eq:B}
\end{align}
so B tries to find the best channel estimate $\bm{h}$ regarding the least-squares of the observation $\bm{y}$ to the prediction $\lr{\bm{h}\ast\bm{x}}$, also referred to as \emph{autoencoder distortion} \cite{caciularu2020unsupervised}. 

Further, with the operator $\real\mlrb{\cdot}$ returning the real part of a complex number, $\lr{\cdot}^\H$ denoting a vector's conjugate-complex and transpose, $\lr{\cdot}^{\mathrm{T}}$ the transpose, and assuming that $\bm{x}$, $\bm{h}$ and $\bm{y}$ are column vectors, C becomes
\begin{align}
	\textrm{C}\mlr{\bm{\theta},\bm{\Phi},\bm{y}} &= \E_{Q}\mlrb{\lr{\bm{y}-\bm{h}\ast\bm{x} }^{\H} \lr{\bm{y}-\bm{h}\ast\bm{x} } } \nonumber \\
	&=  \|\bm{y}\|^2 - 2\real\mlrb{\bm{y}^\H \cdot \E_{Q}\mlrb{\bm{h}\ast\bm{x} } } + \E_{Q}\mlrb{\|\bm{h}\ast\bm{x} \|^2 } \nonumber \\
	&= \|\bm{y}\|^2 - 2 \lr{\bm{y}^\I}^{\mathrm{T}} \textrm{D}^\I - 2 \lr{\bm{y}^\Q}^{\mathrm{T}} \textrm{D}^\Q + \textrm{E} \ , \label{eq:C}
\end{align}
with 
\begin{align*}
	\textrm{D}^\I &= \bm{h}^\I\ast\E_{Q}\mlrb{\bm{x}^\I} - \bm{h}^\Q\ast\E_{Q}\mlrb{\bm{x}^\Q} \ , \\
	\textrm{D}^\Q &= \bm{h}^\I\ast\E_{Q}\mlrb{\bm{x}^\Q} + \bm{h}^\Q\ast\E_{Q}\mlrb{\bm{x}^\I} \ , \\
	\textrm{E} &= \|\textrm{D}^\I\|^2 + \|\textrm{D}^\Q\|^2 +  \\ 
	&\hspace*{0.3em} |\bm{h}|^2 \ast \lr{\E_{Q}\mlrb{ \lr{\bm{x}^\I}^2 } + \E_{Q}\mlrb{\lr{\bm{x}^\Q}^2} - \E_{Q}\mlrb{\bm{x}^\I}^2 - \E_{Q}\mlrb{\bm{x}^\Q}^2} .
\end{align*}
In contrast to \cite{caciularu2020unsupervised}, we keep the vector notation, which helps to identify an efficient implementation, and we also have to compute 
the expectations for any $M$-\ac{QAM} symbol from the VAE-based equalizer's output by ($c\in\{\I,\Q\}$)
\begin{align*}
	\E_{Q}\mlrb{ \bm{x}^{c} } &= \sum_{m=1}^{\sqrt{M}} \bm{q}^{c,A_m}\mlr{\bm{y}} A_m \ , \\
	\E_{Q}\mlrb{ \lr{\bm{x}^{c}}^2 } &= \sum_{m=1}^{\sqrt{M}} \bm{q}^{c,A_m}\mlr{\bm{y}} A_m^2 \ .
\end{align*}

Similarly to \cite{caciularu2020unsupervised}, we find an analytical solution for $\sigma_{\mathrm{w}}^2$ by partially differentiating the loss function and equating it to zero. In fact, A (see \eqref{eq:A}) does not depend on $\sigma_{\mathrm{w}}^2$ and 
\begin{align}
	\frac{\partial \textrm{B}}{\partial \sigma_{\mathrm{w}}^2 } = -N\frac{1}{\sigma_{\mathrm{w}}^2} + \frac{\textrm{C}}{\sigma_{\mathrm{w}}^4} \overset{!}{=} 0 \quad \Rightarrow\quad \sigma_{\mathrm{w}}^2 = \frac{\textrm{C}}{N}\ . \label{eq:sigma}
\end{align}
Hence, inserting \eqref{eq:sigma} into the loss function and omitting all additive constants yields
\begin{align}
	\tilde{\mathcal{L}}\mlr{\bm{\theta},\bm{\Phi},\bm{y}} &:= \textrm{D}_\textrm{KL} \mlr{ \qxy \| P\mlr{\bm{x}}} + N\ln{\textrm{C}}\ . \label{eq:L_short}
\end{align}

Here, we emphasize that the equalizer can also be designed for any integer oversampling factor $N_{\mathrm{os}}$. If the equalizer incorporates downsampling, e.g., by convolution with stride $N_{\mathrm{os}}$, all vectors can be defined accordingly. However, the size of the expectation vectors does not match the size of the observations $\bm{y}$ anymore, so the term C would not be computable. 
Since the loss is summed over all samples, we can simply match the vectors by inserting $\lr{N_{\mathrm{os}}-1}$ zeros between consecutive samples of $\bm{q}^{c,A_m}\mlr{\bm{y}}$. 
\begin{figure} [!t]
	\begin{center}
		\includegraphics{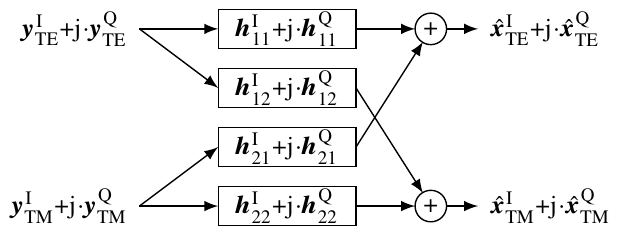}
		\caption{Complex-valued $2\times 2$ \acf{MIMO}-system}   
		\label{fig:2x2}
	\end{center}
\end{figure}

\begin{figure*} [!htb]
	\centering
	\subfloat{
		\includegraphics{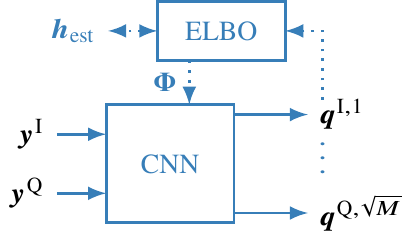}
		 }%
	\subfloat{
		\includegraphics{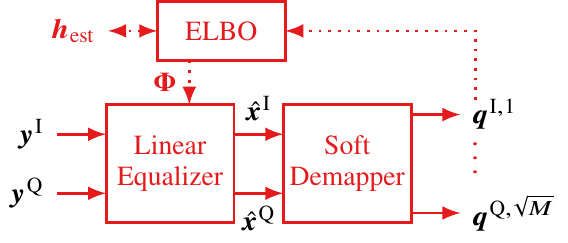}
		}%
	\subfloat{
		\includegraphics{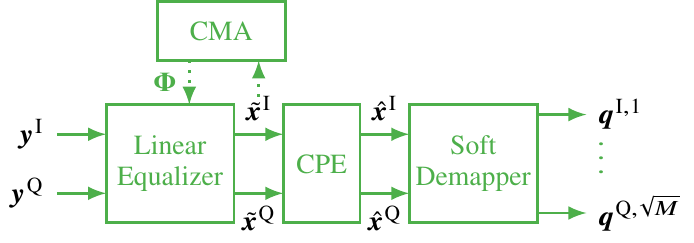}
		}%
	\caption{Receivers considered in this paper for rectangular $M$-QAM formats (with \ac{PCS}): VAE-NN (left), VAE-LE (middle); and reference \ac{CMA} (right).}
	\label{VAE:structures}
\end{figure*}
\subsection{Extension Towards Coherent Optical Communication Systems}
Light is always traveling as combination of two orthogonal polarizations, which can be independently modulated but rotate during propagation through a standard fiber. In combination with other effects like \ac{PMD} and \ac{CD}, the receiver observes a superposition of both polarizations similar to a classical \acf{MIMO} channel with cross-talk. Although we focus on \ac{DP} systems in this work, the proposed VAE-based equalization scheme can be extended towards any kind of \ac{MIMO} system accordingly.  \\

In the \ac{DP} case, the transmitted and received sequences have to be represented by the matrices $\bm{X} = \left[\bm{x}_{\mathrm{TE}} \ \bm{x}_{\mathrm{TM}}\right]$ and $\bm{Y} = \left[\bm{y}_{\mathrm{TE}} \ \bm{y}_{\mathrm{TM}}\right]$, respectively, where the indices denote the transversal electric (TE) and transversal magnetic (TM) polarizations of the laser beam.

Considering the orthogonality, we can model 
\begin{align*}
	Q_{\bm{\Phi}}\mlr{\bm{X}\mid \bm{Y}} &= Q_{\bm{\Phi}}\mlr{\bm{x}_{\mathrm{TE}}\mid \bm{Y}} \cdot Q_{\bm{\Phi}}\mlr{\bm{x}_{\mathrm{TM}}\mid \bm{Y}} \ , \\
	p_{\bm{\theta}}\mlr{\bm{Y}\mid \bm{X}} 
	&= p_{\bm{\theta}}\mlr{\bm{y}_{\mathrm{TE}}\mid \bm{X}} \cdot p_{\bm{\theta}}\mlr{\bm{y}_{\mathrm{TM}}\mid \bm{X}} \ , \\
	P\mlr{\bm{X}} &= P\mlr{\bm{x}_{\mathrm{TE}}} \cdot P\mlr{\bm{x}_{\mathrm{TM}}} \ ,
\end{align*}
and derive the loss function similarly to the \ac{AWGN} channel case, which yields 
\begin{align}
	\tilde{\mathcal{L}}\mlr{\bm{\theta},\bm{\Phi},\bm{Y}} &= \tilde{\mathcal{L}}_{\mathrm{TE}}\mlr{\bm{\theta},\bm{\Phi},\bm{Y}} + \tilde{\mathcal{L}}_{\mathrm{TM}}\mlr{\bm{\theta},\bm{\Phi},\bm{Y}}\ . \label{eq:L_DP_short}
\end{align}
In principal, the losses per polarization are calculated similarly to the proposed case for a single channel. However, in order to incorporate cross-talk between the polarizations, the mean of the likelihood's circular-symmetric normal distribution is not anymore $\lr{\bm{h}\ast \bm{x}}$ (see \eqref{eq:pyx_model}), but depends on the superposition of both polarizations. In this work, we implement it as a complex-valued $2\times 2$ \ac{MIMO}-system as depicted in Fig.~\ref{fig:2x2}, which is based on a physical model~\cite{raheli1991synchronous, Savory_AS_2010digital}. Alternatively, a real-valued $4\times 4$ system can be implemented, which has additional degrees of freedom to compensate transceiver impairments.

\section{Realization of the Equalizer} \label{sec:realization}
Commonly in the machine learning community, the \ac{VAE}'s \emph{encoder} $p_{\bm{\theta}}\mlr{\bm{y}|\bm{x}}$ and the \emph{decoder} $\qxy$ are implemented as \acp{NN} with the parameters $\bm{\theta}$ and $\bm{\Phi}$, but this is no requirement. In the application of the \ac{VAE} concept to communications, the transmission channel is forming the encoder, while the decoder can be either an \ac{NN}, as in in \cite{caciularu2020unsupervised}, or an \ac{FIR} filter system with a soft-demapping block, as proposed in this work. \\

In comparison to the \ac{FIR} filter system, the \ac{NN}
\begin{itemize}
	\item carries out a classification task and, hence, combines equalizer and demapper,
	\item is potentially capable of compensating non-linearities,
	\item requires more learnable parameters since its dimensionality depends on the modulation order,
	\item comprises more hyperparameters which have to be tuned,
	\item provides no access to the output constellation (since it only outputs the approximations $\bm{q}^{c,A_m}\mlr{\bm{y}}$), which may prevent the inclusion into state-of-the-art \ac{DSP} chains.
\end{itemize} 

In Fig.~\ref{VAE:structures}, we show the block diagrams of the investigated adaptive equalizers. The VAE-NN employs a \ac{CNN} with two one-dimensional convolutional layers as in \cite{caciularu2020unsupervised}. Adaptions are necessary to the final layer, namely exchanging the sigmoid with a softmax, to transform the multilevel output into probabilities, and we apply an \ac{ELU} instead of a softsign activation to the first layer, which provides better results. If applicable, a stride in the final layer downsamples the output. We found that the second layer's kernel size can be fixed to a small value (3 to 5), while the first layer's kernel size remains a hyperparameter. Precisely, the first layer's kernel size and the length of the estimated channel impulse response can be any odd integer (to ensure symmetry around a major central tap), which we optimized during our simulations. Typical values for both have been around 29 (for $N_{\mathrm{os}}=2$) and 11 (for $N_{\mathrm{os}}=1$). 
All equalizers and the simulation environment are implemented in Python with the PyTorch library. \\

\subsection{The Proposed VAE-LE Scheme}
We further propose the VAE-LE scheme, which is based on a classical $2\times 2$ butterfly equalizer system with complex-valued \ac{FIR} filters as depicted in Fig.~\ref{fig:2x2}. It uses the same filter system with $F$ taps per filter as the reference \ac{CMA}, which allows the integration into state-of-the-art \ac{DSP} chains \cite{Savory_AS_2010digital}. Note, however, that the computation of the cost function may not be as simple as for the \ac{CMA} and needs to be adapted as it requires a soft-demapper output and backpropagation through the latter as well as possibly some further \ac{DSP} blocks. Future work may take into account backpropagation through consecutive, possibly non-differentiable \ac{DSP} blocks (see, e.g., \cite{rodeOFC} for an example) or the complexity reduction of the update algorithm.
Precisely, the VAE-LE uses one filter system for the equalization, where its weights are $\Phi$ in the derivation above, and a second similar filter system as channel model for the estimation. Both parameter sets are independent, but they are trained simultaneously.  
We also initialize both similarly, i.e., we only initialize the real part of the filters $\bm{h}_{ii}$ with a $1$ at the center tap, while all other taps (including the imaginary parts) are zero (Dirac initialization). Similarly to the \ac{CMA}, if used along with soft-decision \ac{FEC}, the VAE-LE also requires a soft-demapping block to transform the constellation output into the variational approximations.
\begin{figure} [!tb]
	\centering
	\includegraphics{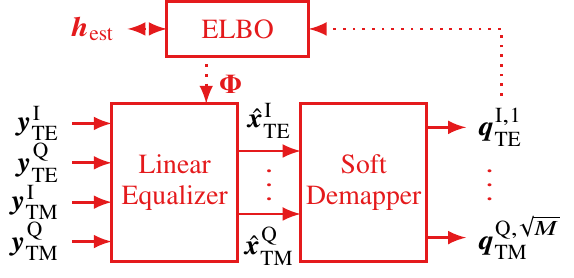}%
	\caption{Structure of the VAE-LE for \ac{DP} systems---the other equalizers are implemented accordingly.}
	\label{VAE:structure_DP}
\end{figure}
The implemented structure of the VAE-LE for a \ac{DP} system is displayed in Fig.~\ref{VAE:structure_DP}, which is representative for the implementation of the other equalizers. 

We assume the general case of squared-$M$-\ac{QAM} transmission, where the modulation symbols' prior \ac{pmf} follows a Maxwell-Boltzmann distribution with normalization constant $C_\nu$ and shaping parameter $\nu\geq0$, i.e.,
\begin{align*}
	p_{\mathrm{MB}}\mlr{x^{c}_i} &= C_\nu \e^{- \nu\lr{x^{c}_i}^2 } , \quad c\in\{\I,\Q\} \ ,
\end{align*}
with $p_{\mathrm{MB}}\mlr{x_i} = p_{\mathrm{MB}}\big(x^\I_i\big)p_{\mathrm{MB}}\big(x^\Q_i\big)$.
For $\nu =0$, $p_{\mathrm{MB}}\mlr{x^{c}_i}$ becomes a uniform distribution. The Maxwell-Boltzmann distribution is the preferred distribution for \ac{PCS}~\cite{bocherer2015bandwidth,BuchaliJLT}, so we can cover both the case of uniform and \ac{PCS}-\ac{QAM} transmission by considering this prior \ac{pmf}. 
We characterize the different formats using the constellation entropy 
\begin{align*}
	\mathcal{H} &= -\E_{p_{\mathrm{MB}}}\{\log_2{p_{\mathrm{MB}}\mlr{x_i} } \} \overset{\lr{i}}{\leq} \log_2{M} \ ,
\end{align*}
with equality in $\lr{i}$ if $\nu=0$ (uniform distribution).
Since optimum demapping is based on the \ac{MAP} criterion~\cite[Ch.~4.1]{proakis2008digital}, we want to find
\begin{align*}
	x_{\mathrm{dec},i}^c &= \arg \max_{x^c_i\in\mathcal{X}^c} p\mlr{x^c_i|y^c_i}\\
	&= \arg \max_{x^c_i\in\mathcal{X}^c} p\mlr{y^c_i|x^c_i} p\mlr{x^c_i} \\
	&= \arg \max_{x^c_i\in\mathcal{X}^c} \lr{ - \frac{ \lr{y^c_i-x^c_i}^2 }{ 2\sigw^2} - \nu \lr{x^{c}_i}^2}
\end{align*}
for a Gaussian likelihood \cite{cho2019probabilistic}. Translating this to \emph{soft}-demapping and adapting it to the $M$-\ac{QAM} case with $x^c_i\in \mathcal{A} = \{A_1, \ \ldots  A_{\sqrt{M}}\}$ yields
\begin{align*}
	\bm{q}_i^{c}\mlr{\bm{y}} &= \mathrm{softmax}\mlr{ -\frac{\lr{\hat{x}_i^{c}\cdot \bm{1}_{\sqrt{M}}-\bm{A}}^2}{2\sigw^2} - \nu\bm{A}_{\mathrm{sq}} }\ ,
\end{align*}
where $\bm{q}_i^{c}\mlr{\bm{y}}=\lr{q_i^{c,A_1}, \ \ldots q_i^{c,A_{\sqrt{M}}}}$, $\hat{x}_i^{c}$ is the equalizer's output constellation (real or imaginary part) at time instant $i$, $\bm{A} = \lr{A_1, \ \ldots A_{\sqrt{M}}}$, $\bm{A}_{\mathrm{sq}} = \lr{A_1^2, \ \ldots A_{\sqrt{M}}^2}$, $\bm{1}_{\sqrt{M}}$ is a vector of $\sqrt{M}$ ones, and $\mathrm{softmax}\mlr{\bm{z}} = \exp\mlr{\bm{z}} /\lr{\sum_{l=1}^{L} \exp\mlr{z_i} }$ with $\bm{z}\in\mathbb{R}^L$.

\subsection{Parameter Update Schemes}
Next, we show how to adapt the \ac{VAE}'s training procedure to resemble the online-update of classical gradient-descent-based equalizers (such as the \ac{CMA}) and to enable tracking of time-varying channels.
Instead of separating the dataset into a training, test, and validation set as in classical supervised machine learning systems, we can directly train on the same data which we evaluate. 
Therefore, we continuously buffer the received data stream, slice it into consecutive mini-batches (of length $N_{\mathrm{B}} \cdot N_{\mathrm{os}}$ samples) and feed them to the equalizer. 
Then, an Adam optimizer \cite{kingma2014adam} constantly updates the weights after each mini-batch. For the VAE-LE, the equalizer outputs all $N_{\mathrm{B}}$ equalized symbols and the soft-demapper translates them into the corresponding variational approximations, while the VAE-NN's \ac{CNN} directly outputs the $N_{\mathrm{B}}$ aproximated probabilities.

For both, we start the slicing of the next batch at the end of the former, so there are consecutive slices without any gap or overlapping in between. However, there is no requirement for having no overlapping, so we can also start the next slice only $N_{\mathrm{flex}}\cdot N_{\mathrm{os}}$ (instead of $N_{\mathrm{B}} \cdot N_{\mathrm{os}}$) samples after the start of the former slice and, thus, reduce the equalizer output from $N_{\mathrm{B}}$ to $N_{\mathrm{flex}}$ symbols.
This results in an overlap of $\lr{N_{\mathrm{B}}-N_{\mathrm{flex}}}\cdot N_{\mathrm{os}}$ samples. 
Hence, each sample is considered for $n = \left\lfloor N_{\mathrm{B}}/N_{\mathrm{flex}}\right\rfloor$ consecutive update steps and, if $\lr{ N_{\mathrm{B}}/N_{\mathrm{flex}}}$ is no integer, some samples are also considered for a further update step. 
This boosts convergence speed by sacrificing computational complexity due to more frequent updates. We call this generalized implementation \emph{VAEflex} and introduce it for the evaluation of the time-varying channel. 
Throughout this work, we focus on adaptive channel equalization assuming an infinitely long random data sequence. Hence, we do not have to worry about overfitting during training for a specific channel and data sequence, although time-varying effects require a continuous re-adaptation of the filters.

The batch-wise update incorporates an implicit averaging of the loss over $N_{\mathrm{B}}\cdot N_{\mathrm{os}}$ samples, while the \ac{CMA} updates the filter taps after each processed symbol by gradient descent as in its standard implementation~\cite{godard, johnson1998blind}. 
We observe that the VAE-LE with batch-wise training, which only updates every $N_{\mathrm{B}}$ symbols, had a significantly shorter computation time on a standard laptop's CPU as the \ac{CMA} with its symbol-wise update.

The constant modulus criterion is not phase sensitive, so the \ac{CMA} can only equalize the amplitude and requires a consecutive \ac{CPE} stage, which we implement using the Viterbi-Viterbi algorithm \cite{viterbi1983nonlinear} with averaging over 501 symbols. This \ac{CPE} stage has, to the best of our knowledge, not been considered in \cite{caciularu2018blind,caciularu2020unsupervised} (or was insufficient), which could explain the severe observed \ac{SER} penalties in their QPSK simulations. In fact, we would expect significantly better results for a \ac{PSK}-transmission over an \ac{AWGN} channel, for which the \ac{CMA} is well suited, especially if sufficiently long data sequences are used.

\section{Results} \label{sec:results}
We evaluated the proposed equalizers both for simulations of a simple \ac{AWGN} channel with \ac{ISI} and an optical \ac{DP} transmission system as well as a time-varying channel as introduced in Sec.~\ref{sec:sys_model}.
The source code is available online~\cite{source_code}.

\subsection{Simulation Environment}
\begin{figure} [!tb]
	\begin{center}	
    	\subfloat{
    		\includegraphics{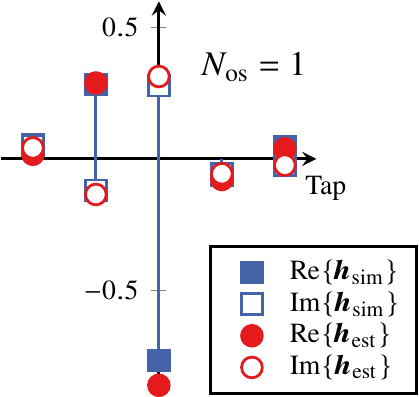}
		}%
    	\hspace*{-22pt}
    	\subfloat{
    		\includegraphics{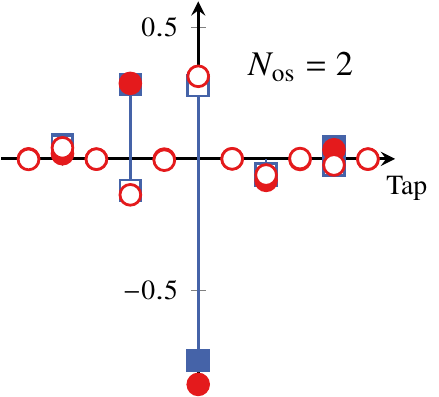}
		}%
		\caption{Simulated channel \ac{IP} $\bm{h}_{\mathrm{sim}}$ (similar to \cite{caciularu2018blind, caciularu2020unsupervised}) and estimations $\bm{h}_{\mathrm{est}}$ by the VAE-LE at $\SI{20}{dB}$ for $N_{\mathrm{os}}=1$ and $N_{\mathrm{os}}=2$ \ac{sps} (without pulse shaping).}
		\label{Channel}
	\end{center}
\end{figure}

Unless stated differently, our transmitter model consists of a source, which outputs a random sequence of $M$-\ac{QAM} symbols, and a \ac{RRC} pulse-shaping filter with roll-off $\alpha=0.1$. We use an oversampling factor of $N_{\mathrm{os}}=2$ \ac{sps} throughout our simulations and incorporate downsampling to the equalizer. 
Since we omit matched filtering (see \cite[Ch.~9]{proakis2008digital}) but expect the equalizers to learn it, the receiver faces \ac{ISI} from the \ac{RRC} pulse-shaping additionally to the simulated channel. 

In the \ac{AWGN} channel, we convolve the transmitted sequence with the complex-valued channel \ac{IP} 
\begin{align*}
	\bm{h}_{\mathrm{sim}} &= [0.055 + \j 0.05, \ 0.283 - \j 0.120, \ -0.768 + \j 0.279, \\
	&\hspace*{30mm} -0.064 -\j 0.058, \ 0.047 -\j 0.023] 
\end{align*}
already used in \cite{caciularu2018blind, caciularu2020unsupervised} (also shown in Fig~\ref{Channel}),
which we oversample by inserting $\lr{N_{\mathrm{os}}-1}$ zeros between consecutive samples and interpolate by convolving it with the \ac{RRC} pulse.
We add real-valued \ac{AWGN} with a variance of $\sigw^2/2$ on both real and imaginary part and take the oversampling into account for the \ac{SNR} calculation.

Monte-Carlo simulations of an \ac{AWGN} channel without \ac{ISI} provide a baseline \ac{SER} per \ac{SNR} and modulation format, which we denote by ``No ISI'' in the result plots.

Our optical \ac{DP} transmission model follows \cite{DEq_CD_PMD_ip2007digital} and \cite{Savory_DF_2008digital} and includes a static (input and output) IQ-phase-shift $\varphi_{\mathrm{IQ}}$ and a static rotation of the reference polarization to the fiber's \ac{PSP}, called HV-phase-shift $\gamma_{\mathrm{hv}}$. 
Additionally, we simulate both first-order \ac{PMD} caused by the differential group delay $\tau_{\mathrm{pmd}}=D_{\mathrm{pmd}}\sqrt{L_{\mathrm{pmd}}}$ between the \acp{PSP} over a fiber length $L_{\mathrm{pmd}}$,
and residual \ac{CD}, which is defined by the fiber's group velocity dispersion (GVD) parameter $\beta_{\mathrm{cd}}$ \cite[Ch.~2.3]{agrawal2010fiber} times the uncompensated fiber length $L_{\mathrm{cd}}$. 
We display all parameter values in Table~\ref{table:sim_param}. 

\begin{table}[!tb]
	\centering
	\caption{Simulation parameters}
	\begin{tabular}{c c c c c}
		\toprule
		{Parameter}&  {Value}& \hspace{7pt} & {Parameter}&  {Value}\\
		\midrule
		$\gamma_{\mathrm{hv}}$&  $0.1\pi$&  \hspace{7pt} &$\varphi_{\mathrm{IQ}}$&  $0.01\pi$\\
		$D_{\mathrm{pmd}}$&  $\SI{0.1}{\pico\s\per\sqrt{\kilo\m}}$ \cite{DEq_CD_PMD_ip2007digital}&  \hspace{7pt} &$L_{\mathrm{pmd}}$&  $\SI{1000}{\kilo\m}$\\
		$\beta_{\mathrm{cd}}$\footnotemark[1]& $\SI{-26}{\pico\s^2\per\kilo\m}$ & \hspace{7pt} &$L_{\mathrm{cd}}$& $\SI{1}{\kilo\m}$\\
		\bottomrule 
		\multicolumn{5}{l}{\vspace*{-4pt} }\\
		\multicolumn{5}{l}{\footnotemark[1] equals $D_{\mathrm{cd}} = \SI{20}{\pico\s\per\nano\m\per\kilo\m}$ at $\lambda=\SI{1550}{\nano\m}$ (see \cite[Eq.~(2.3.5)]{agrawal2010fiber})}
	\end{tabular}
	\label{table:sim_param}
\end{table}
By assuming---similarly to \cite{DEq_CD_PMD_ip2007digital}---that all non-linear effects and phase noise are either negligible or compensated, and that we transmit a complex-valued \ac{DP} signal $\bm{x} = \left[x_{\mathrm{TE}}\mlr{t} \ x_{\mathrm{TM}}\mlr{t}\right]$, we can model the fiber by the linear time-invariant frequency domain channel matrix
\begin{align*}
	\bm{H}\lr{f} &= \bm{R}^{\mathrm{T}} \begin{pmatrix} \e^{\j\pi \tau_{\mathrm{pmd}}  f } & 0 \\ 0 & \e^{-\j\pi \tau_{\mathrm{pmd}}  f } \end{pmatrix} \bm{R} \cdot \e^{-\j2\pi^2 \beta_{\mathrm{cd}} L_{\mathrm{fiber}}  f^2 } 
\end{align*}
where $\bm{R} = \e^{-\j\varphi_{\mathrm{IQ}}}\begin{pmatrix} \cos{\gamma_{\mathrm{hv}}} & \sin{\gamma_{\mathrm{hv}}} \\ -\sin{\gamma_{\mathrm{hv}}} & \cos{\gamma_{\mathrm{hv}}} \end{pmatrix}$.
Again, we add \ac{AWGN} on both polarizations and I/Q-components with a variance of $\sigw^2/2$ each.
To simulate a time-varying channel, we change the HV-phase-shift after each frame of $N_{\mathrm{frame}} = 10,000$ symbols. 
Precisely,
we extend 
$\tilde{\gamma}_{\mathrm{hv}} = \gamma_{\mathrm{hv}} + \Delta \gamma_{\mathrm{hv}} \cdot kT_{\mathrm{frame}}$ 
with the deviation of the HV-shift $\Delta \gamma_{\mathrm{hv}}$,
the frame duration $T_{\mathrm{frame}} = N_{\mathrm{frame}} / R_{\mathrm{S}}$, the symbol rate $R_{\mathrm{S}}$ and the frame-index $k$ as an integer indexing the discrete time (in multiples of the frame duration). It should be highlighted that we apply deviations frame-wise, i.e., we neglect deviations within each frame of $N_{\mathrm{frame}}$ symbols. 

\begin{figure} [!tb] 
	\begin{center}
		\includegraphics{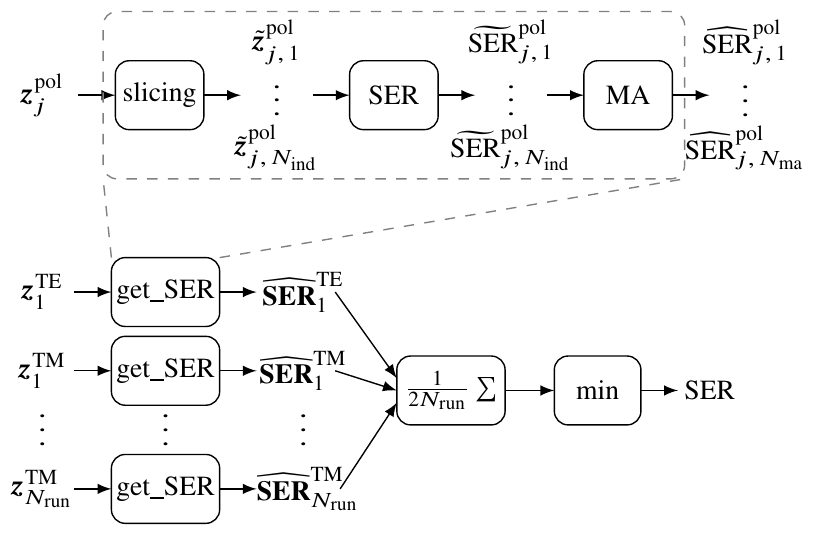}		
		\caption{Block diagram sketching the result averaging---$\tilde{\bm{z}}^{\mathrm{pol}}_{j,\,k}\in \mathbb{C}^{N_\mathrm{frame}}$ 
			are the slices (frames) of the vector containing the corresponding equalizer output with $\mathrm{pol}\in\{\TE, \TM\}$, MA is a moving average filter with length $F_{\mathrm{ma}} = 10$, so $N_{\mathrm{ma}} = (N_{\mathrm{ind}}-F_{\mathrm{ma}}+1)$, and min is an operator which returns a vector's element with the minimum value.}
		\label{results:averaging}
	\end{center}
\end{figure}

For evaluation, we chose the averaging scheme sketched in Fig.~\ref{results:averaging}, which might not seem to be straight-forward at first glance but fits to our needs, i.e., it returns a reliable \ac{SER} estimate to compare different hyperparameter settings but also provides insights into the convergence behavior. In particular, the analysis of the average convergence behavior requires a certain temporal averaging, but the final analysis requires also an averaging over multiple execution runs, since the behavior could vary between different runs.

To get the desired insights, we first have to slice the equalizer output $\bm{z}^{\mathrm{pol}}_j$ (per simulation run $j$ and per polarization $\mathrm{pol}\in\{\TE, \TM\}$)  into $N_{\mathrm{ind}} = 170$ consecutive frames $\tilde{\bm{z}}^{\mathrm{pol}}_{j,\,k}\in \mathbb{C}^{N_\mathrm{frame}}$ to allow the evaluation at different time steps, respectively frame-indices $k = 1\,\ldots\,N_{\mathrm{ind}}$.
Then, we estimate each frame's scalar $\widetilde{\mathrm{SER}}^{\mathrm{pol}}_{j,\,k}$ after hard decision taking into account the prior distribution of the constellation symbols~\cite{cho2019probabilistic}. Since we perform blind equalization, we also need to compensate for possible time-shifts, I/Q-flips and phase-rotations (in multiples of $\frac{\pi}{4}$). Furthermore, we discard symbols that may be incorrect due to boundary effects (but having at least 8,000 symbols per frame remaining for evaluation---hence, we can evaluate the potential performance of each algorithm).
Eventually, we perform a moving average with filter length $F_{\mathrm{ma}} = 10$ and get a sequence of $N_{\mathrm{ma}} = N_{\mathrm{ind}}-F_{\mathrm{ma}}+1$ estimates $\widehat{\mathrm{SER}}^{\mathrm{pol}}_{j,\,k}$, each being evaluated on approximately $80,000\, \ldots \,100,000$ symbols per frame-index $k$ (per polarization and per run). This allows the analysis of the equalizers' convergence for each polarization and run independently.

For further evaluations, we need one reliable scalar \ac{SER} estimate for each algorithm and hyperparater setting. Thus, we carry out $N_{\mathrm{run}} = 10$ (unless stated otherwise) independent simulation runs and average the results over all successful runs, i.e., runs where the \ac{SER} drops below a pre-defined threshold of $0.3$, by taking the element-wise mean of the vectors $\widehat{\mathbf{SER}}^{\mathrm{pol}}_{j}\in\mathbb{R}^{N_\mathrm{ma}}$, which contain the estimates per run and polarization after moving average. If unsuccessful runs occurred, we display the amount by a small number next to the corresponding data point in the result figures. Finally, we get a reliable scalar \ac{SER} estimate by taking the element with the minimum value from the vector containing the averaged estimates for all remaining $N_\mathrm{ma}$ frame-indices.
We chose the minimum over the mean to display the best (averaged) performance of the equalizers, since we already have a sufficient averaging (the mean of $N_{\mathrm{pol}}\cdot N_{\mathrm{run}}$ estimates where each is averaged over at least 80,000 symbols), but can prevent distortions from potential outliers. In fact, we did not observe any outliers in all our simulations, so we conjecture that the estimates' variance is small and the difference between taking the minimum and the mean (after convergence) is insignificant. 

\begin{figure*} [!htbp]
	\begin{center}
		\subfloat{
			\includegraphics{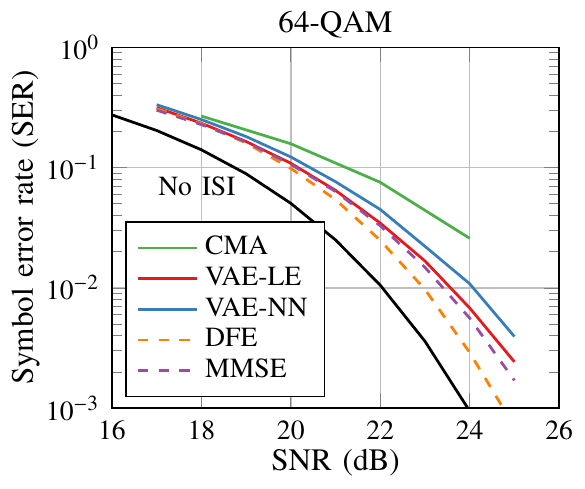}
			} %
		\hspace*{-10pt}
		\subfloat{
			\includegraphics{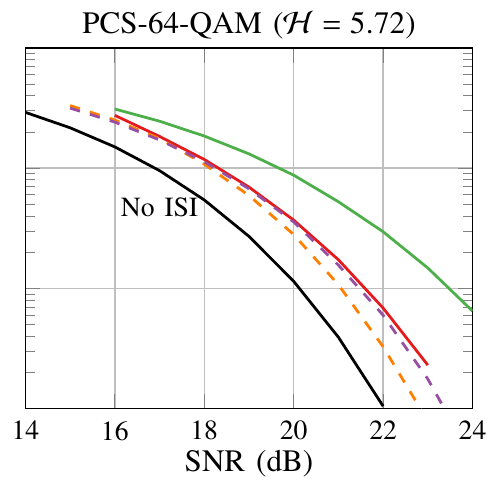}
			}%
		\hspace*{-10pt}
		\subfloat{
			\includegraphics{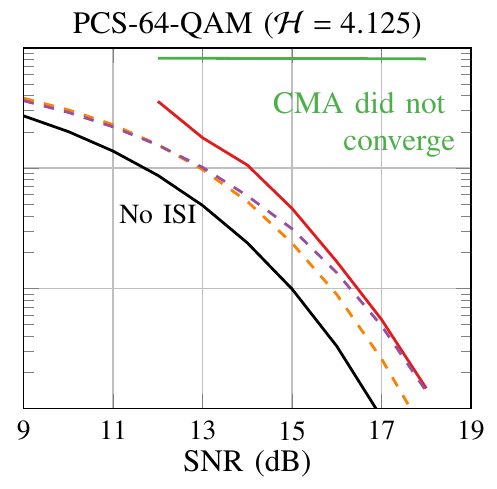}
			}%
		\caption{Results for uniform and \ac{PCS}-64-\ac{QAM} through an \ac{AWGN} channel with \ac{ISI} and without learning rate reduction---see also~\cite{lauinger2021blind}; the simulations for the \emph{non-blind} DFE and MMSE are without pulse-shaping at $N_{\mathrm{os}}=1$~\ac{sps}.}
		\label{Results:AWGN}
	\end{center}
\end{figure*}

\begin{figure} [!htbp]
	\begin{center}
		\includegraphics{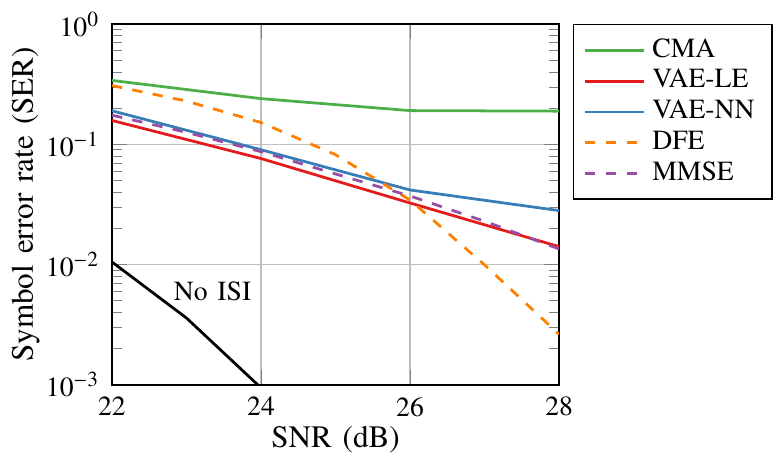}
		\caption{Results for uniform 64-\ac{QAM} through an \ac{AWGN} channel with \ac{ISI} and without learning rate reduction---see also~\cite{lauinger2021blind}: $\bm{h}_{\mathrm{sim},2} = [0.055 + \j 0.017, \ -1.345 - \j 0.452, \ 1.007 + \j 1.152, \ 0.348 +\j 0.315]$; the simulations for the \emph{non-blind} DFE and MMSE are without pulse-shaping at $N_{\mathrm{os}}=1$~\ac{sps}.}
		\label{Results:AWGN_h2}
	\end{center}
\end{figure}  

Furthermore, we use a learning rate scheduler for the optical \ac{DP} channel which halves the learning rate $\elr$ of each equalizer after every frame-index $k$ being an integer-multiple of 20, so the given $\elr$ represents the initial value.

\subsection{Simple \ac{AWGN} Channel with \ac{ISI}}
Figure~\ref{Results:AWGN} depicts the proposed equalizers' performance for both uniform and \ac{PCS}-64-\ac{QAM} at $N_{\mathrm{os}}=\SI{2}{sps}$. 
We tuned the lengths of the filters $F$, the batches $N_{\mathrm{B}}$, the kernels, and the learning rate.
Later, for the \ac{DP} simulations, we will show how the hyperparameters influence the equalizers' performance.
Besides the baseline ``No ISI''-curve and the \ac{CMA} as reference, we also evaluated two \emph{non-blind} equalizers at $N_{\mathrm{os}}=\SI{1}{sps}$ without pulse-shaping, namely the non-linear \ac{DFE} \cite[Ch.~9.5]{proakis2008digital} (with 10 taps each for both feed-forward and feedback filters) and the linear \ac{MMSE} equalizer \cite[Ch.~9.4]{proakis2008digital} (with 20 taps). 

For uniform 64-QAM, the CMA converges with a significant penalty to the MMSE, while both \ac{VAE}-based equalizers operate close to the \ac{MMSE}. For a linear channel, the VAE-LE outperforms the VAE-NN, which might originate from a closer bounding of the family $\mathcal{Q}$ around the optimum $\hat{Q}\mlr{\bm{x}|\bm{y}} \in \mathcal{Q}$. We found qualitatively similar results for other \ac{ISI} channels as well, e.g., in Fig.~\ref{Results:AWGN_h2} for channel $\bm{h}_2$ from~\cite{caciularu2020unsupervised}. For the sake of computational complexity, we restrict ourselves in the following to the VAE-LE.

As already forecasted in the early 1990s \cite{chan1990stationary, zervas1991effects}, the CMA fails to converge for \ac{PCS} formats which approximate a Gaussian prior; however, the VAE-LE even reaches the non-blind MMSE's performance for higher \acp{SNR}, although we used the same soft-demapper as for the uniform \ac{QAM} in these \ac{AWGN} channel simulations (instead of the optimal \ac{PCS}-adapted version as introduced in Sec.~\ref{sec:realization}).
\begin{figure} [!tbp]
	\begin{center}
		\includegraphics{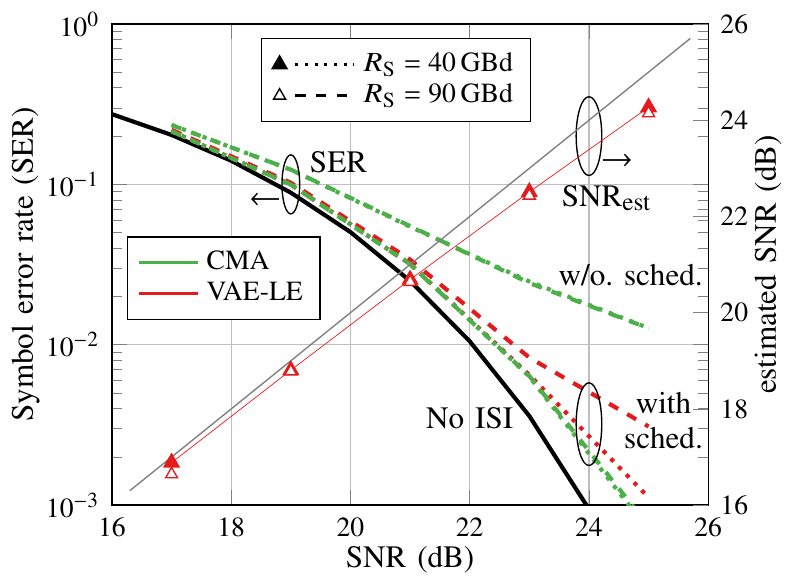}
		\caption{Performance of the VAE-LE and the CMA for uniform \ac{DP}-64-\ac{QAM} with $\elr=\num{0.5e-3}$, $F=25$ and $N_{\mathrm{B}}=350$: \ac{SER}, left axis, and estimated \ac{SNR}, right axis, vs \ac{SNR} for $R_{\mathrm{S}}=\SI{40}{\giga Bd}$ and $R_{\mathrm{S}}=\SI{90}{\giga Bd}$. The CMA is evaluated with ($\elr=\num{0.5e-3}$) and without (w/o. -- $\elr=\num{5e-5}$) learning rate scheduler.}
		\label{Results:SER_SNR_SNR_DP}
	\end{center}
\end{figure} 
\begin{figure} [!tbp]
	\begin{center}
		\includegraphics{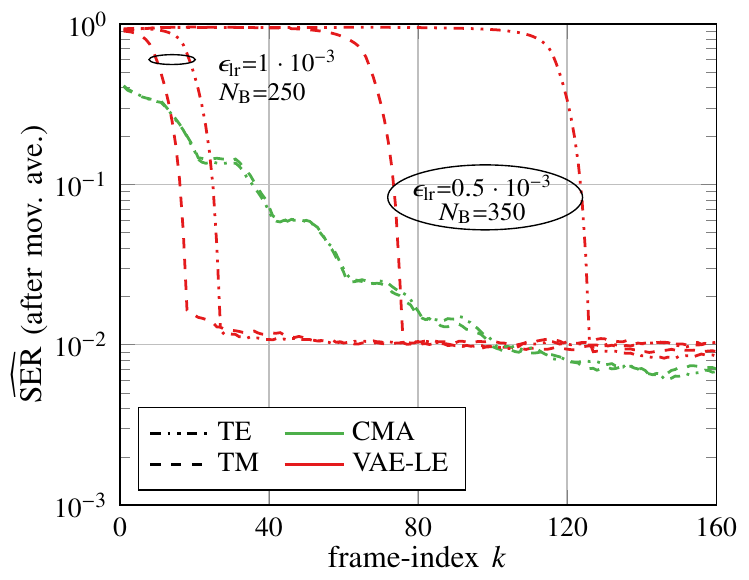}
		\caption{Convergence behavior of the VAE-LE and the CMA ($\elr= \num{0.5e-3}$) for uniform DP-64-\ac{QAM} with $R_{\mathrm{S}}=\SI{90}{\giga Bd}$, $\SI{23}{dB}$ and $F=25$.}
		\label{Results:SER_verlauf}
	\end{center}
\end{figure} 
Furthermore, Fig.~\ref{Channel} demonstrates the VAE-LE's capability of estimating the channel \ac{IP}. We depict the estimate without averaging while processing uniform 64-\ac{QAM} at $\mathrm{SNR} = \SI{20}{dB}$ for 1 and 2 \ac{sps} without pulse-shaping.

\subsection{Optical Dual-polarization Transmission}

\begin{figure*} [!htbp]
	\begin{center}
		\subfloat{
			\includegraphics{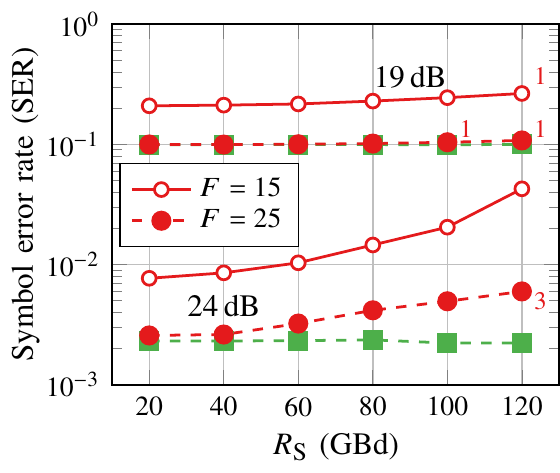}
			} %
		\hspace*{-10pt}
		\subfloat{
			\includegraphics{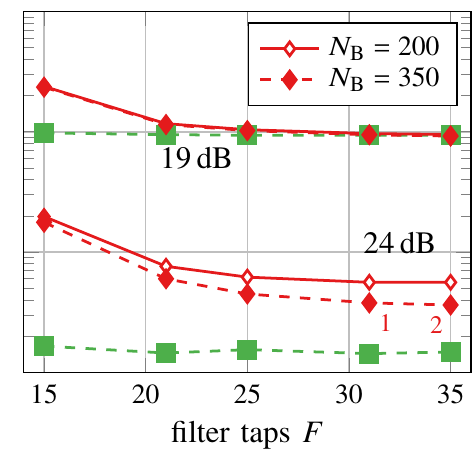}
			}%
		\hspace*{-10pt}
		\subfloat{
			\includegraphics{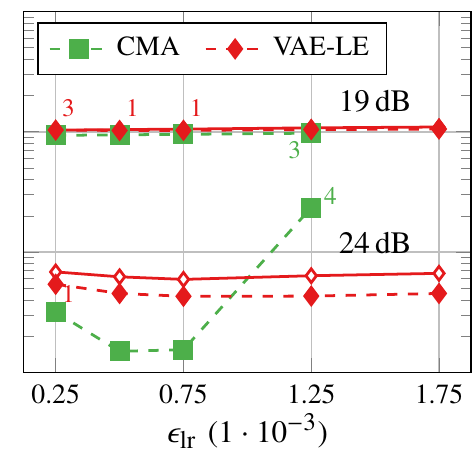}
			}%
		\caption{Hyperparameter analysis of the VAE-LE (\textcolor{ALUColor2}{red, \textbf{------}}) and the CMA (\textcolor{ALUColor1}{green, \textbf{------}}) for uniform \ac{DP}-64-\ac{QAM} with $R_{\mathrm{S}}=\SI{90}{\giga Bd}$ and the default values of $F=25$, $\elr=\num{0.5e-3}$ and $N_{\mathrm{B}}=350$.}
		\label{Results:DP64_details}
	\end{center}
\end{figure*} 

\begin{figure*} [!htbp]
	\begin{center} 
		\subfloat{
			\includegraphics{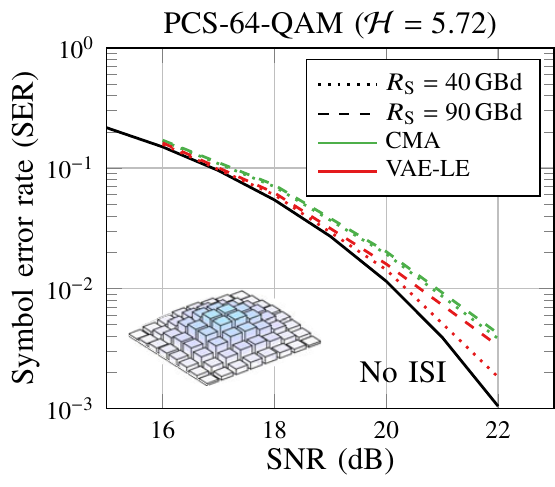}
			}%
		\hspace*{-10pt} 
		\subfloat{
			\includegraphics{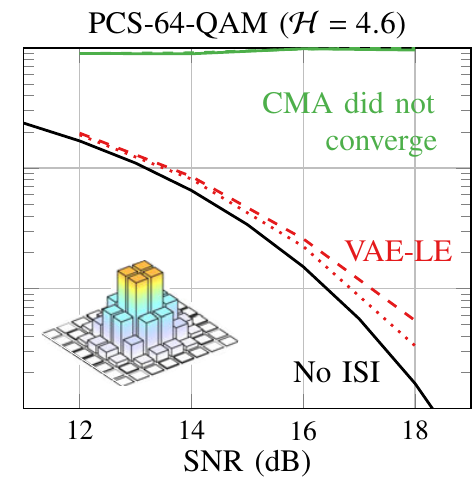}
			}%
		\hspace*{-10pt}
		\subfloat{
			\includegraphics{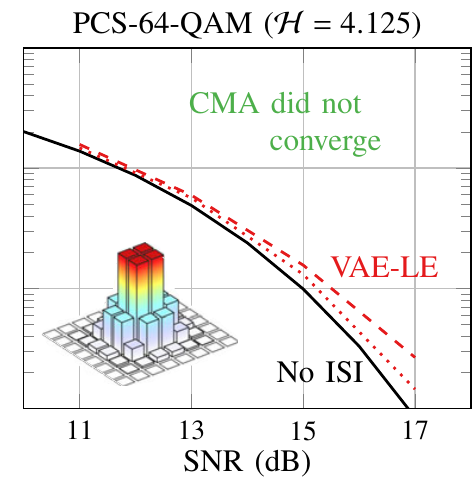}
			}%
		\\
		\vspace*{-10pt}
		\subfloat{
			\includegraphics{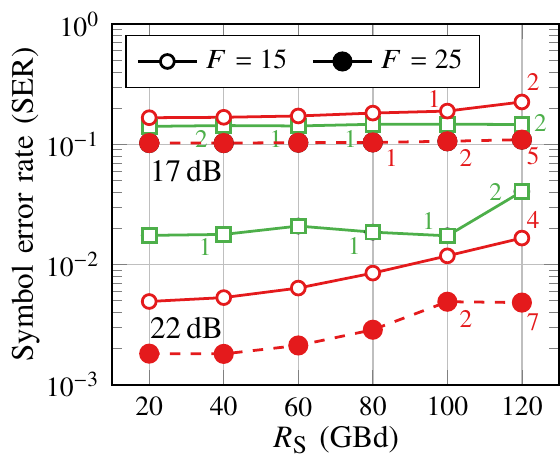}
			}%
		\hspace*{-10pt}
		\subfloat{
			\includegraphics{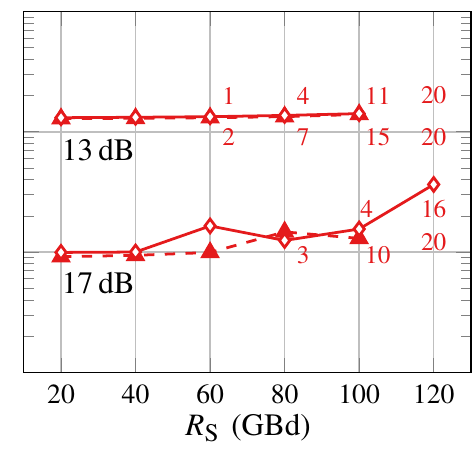}
			}%
		\hspace*{-10pt}
		\subfloat{
			\includegraphics{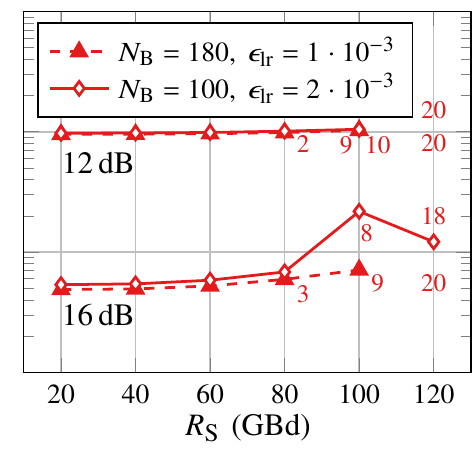}
			}%
		\caption{Results for \ac{DP}-\ac{PCS}-64-\ac{QAM} with VAE-LE (\textcolor{ALUColor2}{red, \textbf{------}}) and CMA (\textcolor{ALUColor1}{green, \textbf{------}}); for $\mathcal{H} = 5.72$: $\elr=\num{0.5e-3}$ and $N_{\mathrm{B}}=350$; for $\mathcal{H} = 4.6$ and $\mathcal{H} = 4.125$: $\elr=\num{1e-3}$ (VAE-LE), $N_{\mathrm{B}}=180$, $F=15$, and $N_{\mathrm{run}}=20$ for better statistics.}
		\label{Results:SER_SNR}
	\end{center}
\end{figure*}

The results for the application of the VAE-LE are depicted in Fig.~\ref{Results:SER_SNR_SNR_DP}.
The right axis shows the estimated \ac{SNR} (averaged similarly to the \ac{SER}) while evaluating \ac{DP}-64-\ac{QAM} at various \acp{SNR}. The VAE-LE always underestimates the \ac{SNR}, but only within a fraction of a decibel (dB). Along with the results of Fig.~\ref{Channel}, this demonstrates the potential for joint communications and sensing, since the estimation of the channel parameters is a valuable byproduct for tracking and evaluating the channel \ac{IP} and \ac{SNR} without interfering communications.

The left axis in Fig.~\ref{Results:SER_SNR_SNR_DP} depicts the corresponding \acp{SER}. Similar to the results in Fig.~\ref{Results:AWGN}, the CMA has a significant penalty without learning rate scheduler. Since the scheduler halves the learning rate after every frame-index $k$ being an integer-multiple of 20, the \ac{CMA}'s performance-convergence trade-off is avoided and it reaches marginally lower \acp{SER} as the VAE-LE. For high symbol rates and high \acp{SNR}, the VAE-LE deviates from the ideal ``No ISI''-curve. In all other cases, both equalizers stay within $\SI{1}{dB}$ of the ``No ISI''-curve and converge to it for low \acp{SNR}.

The equalizers' convergence behavior differs significantly, as depicted in Fig.~\ref{Results:SER_verlauf}, where we display one simulation run's $N_{\mathrm{ma}}$ SER estimates after the moving average filter. It should be noted that
we do not consider a fixed amount of training data but conduct online-learning on the (in theory, infinitely long) received data sequence and that
the frame-index corresponds to discrete time steps during training, i.e., one frame has a duration of $T_\mathrm{frame}$ and corresponds to $N_\mathrm{os}\cdot N_\mathrm{frame}$ samples processed by the equalizer. The CMA converges gradually with a decreasing slope. The distinct ``plateaus'' are caused by the learning rate scheduler. The VAE-LE shows a ``waterfall-like'' curve.
Also, the VAE-LE starts at a significantly higher \ac{SER}. Although the initial learning rate and batch size strongly influence the VAE-LE's time convergence, the curve's waterfall-like shape remains. Interestingly, the CMA seems to optimize both polarizations equally, while the VAE-LE first focuses on one polarization before trailing the second one.

The influence of the main hyperparameters is analyzed in Fig.~\ref{Results:DP64_details}. Interestingly, the symbol rate $R_{\mathrm{S}}$ and the filter length $F$ influence the CMA significantly less than the VAE-LE, which suffers from performance penalties for low $F$. The reason might be that the VAE-LE has to compensate for both amplitude and phase offset while an extra \ac{CPE} compensates the latter for the CMA. Since a high symbol rate mainly increases the \ac{ISI}, it is reasonable that it only influences the VAE-LE for high \ac{SNR}.  

The VAE-LE suffers from convergence issues for high $F$ and high $N_{\mathrm{B}}$ (or $\elr$), but is very stable for small $F$ and small $N_{\mathrm{B}}$ even under high symbol rates. While changes in the initial learning rate affect the \ac{CMA} strongly, the VAE-LE has a negligible penalty over a relatively large range.

\begin{figure} [!tbp]
	\begin{center}
		\includegraphics{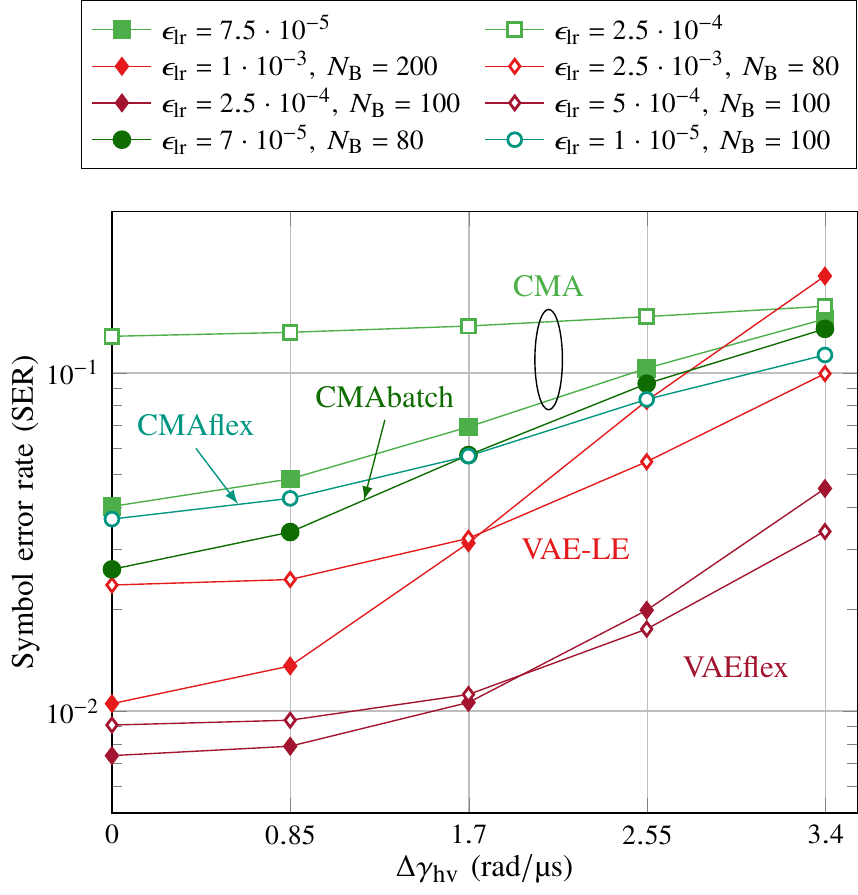} 
		\caption{Results for uniform DP-64-\ac{QAM} transmitted over a time-varying optical \ac{DP} channel with $\mathrm{SNR}=\SI{23}{dB}$, $R_{\mathrm{S}}=\SI{90}{\giga Bd}$ and $F=25$ (here, no learning rate scheduler is used): CMA, CMAbatch, VAE-LE as well as CMAflex and VAEflex with $N_{\mathrm{flex}}=10$. Note that all deviations are simulated frame-wise with $T_{\mathrm{frame}}=\SI{0.111}{\micro\second}$.}
		\label{Results:SER_thetadiff}
	\end{center}
\end{figure} 

We show the results for \ac{PCS} in a \ac{DP} optical channel in Fig.~\ref{Results:SER_SNR}. With the adapted soft-demapping and the learning rate scheduler, the VAE-LE is able to follow the ``No ISI''-curve within $\SI{1}{dB}$ penalty even for high symbol rates and low \ac{SNR}. Lower filter lengths $F$ and batch-sizes $N_{\mathrm{B}}$ as well as higher $\elr$ are necessary to ensure a high probability for convergence, especially for high symbol rates. The VAE-LE is potentially capable of converging for  $R_{\mathrm{S}}>\SI{100}{\giga Bd}$ and strong shaping in our simulation model, but the probability for non-convergence is relatively high for typical working points. Still, the VAE-LE significantly outperforms the CMA for \ac{PCS}, where the latter does not converge at all. 

\subsection{Time-varying Channel}

We analyze the influence of time-dependent channel deviations to the equalizers by a frame-wise increasing HV-shift $\tilde{\gamma}_{\mathrm{hv}} = \gamma_{\mathrm{hv}} + \Delta \gamma_{\mathrm{hv}} \cdot k T_{\mathrm{frame}}$ within the optical \ac{DP} model. Figure~\ref{Results:SER_thetadiff} depicts the \ac{SER} for different slopes $\Delta \gamma_{\mathrm{hv}}$. Since we did not employ the learning rate scheduler in this evaluation, the hyperparameter values differ from the ones used in the time-invariant case. Due to its rather slow convergence speed, the CMA has to operate at rather high learning rates and entails a severe penalty. The wide range of possible learning rates allows the VAE-LE to optimize its working point towards low \acp{SER} for low $\Delta\gamma_{\mathrm{hv}}$ or a high tolerance towards deviations by moderate penalties. The VAEflex with $N_{\mathrm{B}}=100$ and $N_{\mathrm{flex}}=10$ accelerates training significantly, which makes it tolerant towards deviations by still reaching very low \acp{SER}. Although we did not optimize the VAEflex as thoroughly as the other algorithms, it converges until a low (respectively, early) frame-index $k$ at the cost of a higher computational complexity. Hence, an option would be to switch between the VAEflex and the batch-wise VAE during operation by changing $N_{\mathrm{flex}}$ between 1 and $N_{\mathrm{B}}$. 

For comparison, we also implemented a \ac{CMA} with a batch-wise updating scheme as proposed in~\cite{crivelli2014}, which we denote \emph{CMAbatch}. Additionally, we also extended this scheme with a flexible update rule akin to the VAEflex. We denote the resulting equalizer \emph{CMAflex}. Although Fig.~\ref{Results:SER_thetadiff} shows that both the CMAbatch and the CMAflex perform better than the symbol-wise CMA for this time-varying channel and without the learning rate scheduler, the gain is relatively small and both are outperformed by the VAE-based equalizers. Especially the CMAflex performs very similar to the CMAbatch and, in contrast to the VAEflex, the flexible update rule is not able to accelerate training.

\section{Conclusion} \label{sec:conclusion}
In this paper, we proposed the new VAE-LE, a model-based approach with linear butterfly \ac{FIR} filters, which is trained by the \ac{VAE}-based learning paradigm with extensions towards practically relevant optical communication systems with oversampling and \ac{DP}-\ac{PCS}-\ac{QAM} transmission.

For \ac{AWGN} channels with \ac{ISI}, the \emph{blind} VAE-based equalizers can approach the performance of the \emph{non-blind} \ac{MMSE} equalizer for both uniform and \ac{PCS} formats. The proposed VAE-LE outperforms the previously introduced VAE-NN for this linear channel. The VAE-LE equalizer also converges in a dispersive optical \ac{DP} system but shows a negligible penalty to the \ac{CMA} for uniform formats. However, for \ac{PCS} formats where the \ac{CMA} fails to converge without modifications, the VAE-LE still approaches the ideal reference within $\SI{1}{dB}$. 

The VAE-LE's rapid convergence behavior 
is advantageous for time-varying channels, where the gradually converging \ac{CMA} performs significantly worse. Our proposed VAEflex update scheme with flexible step length is a powerful alternative if convergence speed is a key factor.
Additionally, we have shown that the VAE-LE is able to estimate both the communication channel taps and the noise variance very well, which can be an enabler for joint communications and sensing. 

While we focused on linear channels in this work, the extension towards nonlinear impairments might be a possible direction of future research.

\section*{Acknowledgment}
The authors would like to thank Luca Schmid for providing the code for the DFE and MMSE equalizers.

\ifCLASSOPTIONcaptionsoff
  \newpage
\fi

\end{document}